\documentclass[12pt]{article}

\usepackage{scicite}

\usepackage{times}

\usepackage{graphicx}
\usepackage{siunitx}
\usepackage{color,soul}

\newenvironment{sciabstract}{%
\begin{quote} \bf}
{\end{quote}}

\title{Ultrafast response of spontaneous photovoltaic effect in 3R-MoS\textsubscript{2}-based heterostructures}

\author
{Jingda Wu,$^{1,2,\dagger}$ Dongyang Yang,$^{1,2,\dagger}$ Jing Liang, $^{1,2,\dagger}$ Max Werner,$^{1,2}$\\ 
Evgeny Ostroumov,$^{1,2}$ Yunhuan Xiao,$^{1,2}$ Kenji Watanabe,$^{3}$\\
 Takashi Taniguchi,$^{4}$ Jerry I. Dadap,$^{1,2}$ David Jones,$^{1,2}$ Ziliang Ye$^{1,2\ast}$\\
\\
\normalsize{$^{1}$Department of Physics and Astronomy, The University of British Columbia}\\
\normalsize{6224 Agricultural Road, Vancouver, BC V6T 1Z1, Canada}\\
\normalsize{$^{2}$Stewart Blusson Quantum Matter Institute, The University of British Columbia}\\
\normalsize{2355 East Mall, Vancouver, BC V6T 1Z4, Canada}\\
\normalsize{$^{3}$Research Center for Functional Materials, National Institute for Materials Science}\\
\normalsize{1-1 Namiki, Tsukuba 305-0044, Japan}\\
\normalsize{$^{4}$International Center for Materials Nanoarchitectonics, National Institute for Materials Science}\\
\normalsize{1-1 Namiki, Tsukuba 305-0044, Japan}\\
\normalsize{$^\ast$To whom correspondence should be addressed; E-mail:  zlye@phas.ubc.ca.}\\
\normalsize{$^{\dagger}$There authors contributed equally to this work}\\
}

\date{}

\begin{document} 


\baselineskip24pt


\maketitle


\begin{sciabstract}
Rhombohedrally stacked MoS\textsubscript{2} has been shown to exhibit spontaneous polarization down to the bilayer limit and can sustain a strong depolarization field when sandwiched between graphene. Such a field gives rise to a spontaneous photovoltaic effect without needing any p-n junction. In this work, we show the photovoltaic effect has an external quantum efficiency of 10\% for devices with only two atomic layers of MoS\textsubscript{2} at low temperatures, and identify a picosecond-fast photocurrent response, which translates to an intrinsic device bandwidth at $\sim$ 100-GHz level. To this end, we have developed a non-degenerate pump-probe photocurrent spectroscopy technique to deconvolute the thermal and charge-transfer processes, thus successfully revealing the multi-component nature of the photocurrent dynamics. The fast component approaches the limit of the charge-transfer speed at the graphene-MoS\textsubscript{2} interface. The remarkable efficiency and ultrafast photoresponse in the graphene-3R-MoS\textsubscript{2} devices support the use of ferroelectric van der Waals materials for future high-performance optoelectronic applications.
\end{sciabstract}


\section*{Introduction}
When two layers of transition metal dichalcogenides (TMDs) are stacked in parallel, both the inversion symmetry and mirror symmetry are spontaneously broken, leading to an electric polarization along the out-of-plane direction \cite{sung2020broken}. Under an electric field larger than the coercive field, the polarization direction can be flipped as one layer slides relative to the neighboring layer, which has been termed sliding ferroelectricity \cite{wang2022interfacial, vizner2021interfacial, wu2021sliding,weston2022interfacial}. While the artificially stacked bilayers have limited domain sizes due to twist misalignment, homogeneous polarization in a scale as large as the entire flake has been observed in bilayers directly exfoliated from a 3R-MoS\textsubscript{2} bulk crystal \cite{jing2022x}, making it ideal for certain optoelectronic applications. In particular, when the 3R-MoS\textsubscript{2} is sandwiched between graphene \cite{yang2022spontaneous}, the spontaneous polarization gives rise to a photovoltaic effect, known as spontaneous photovoltaic effect \cite{akamatsu2021van}, where photoexcited carriers in MoS\textsubscript{2} transfer asymmetrically to the graphene under a largely unscreened depolarization field. An external quantum efficiency (EQE) of 16\% has been observed in a device with ten-layer 3R-MoS\textsubscript{2}.

The high EQE in the graphene/3R-MoS\textsubscript{2}/graphene heterostructures suggests a potentially ultrafast photocurrent dynamics comparable to the charge-transfer process at the MoS\textsubscript{2}-graphene interface, which lasts approximately one picosecond \cite{hong2014ultrafast, yuan2018photocarrier, Zhang2017, luo2021twist,chen2019highly}. Such an ultrafast photocurrent, if confirmed, can potentially be leveraged for high-speed optical communications. Although the MoS\textsubscript{2} photocarrier dynamics generally can be measured by ultrafast techniques such as optical pump-probe spectroscopy \cite{jin2018ultrafast}, it remains a challenge to probe the intrinsic dynamics of the aforementioned photovoltaic effect, due to the existence of multiple photocurrent contributions. Because the spontaneous polarization naturally induces image charges of opposite polarities at the two electrodes, both the bolometric and the photo-thermoelectric (PTE) effects exist in the device \cite{wu2018broadband, zhang2014ultrahigh,buscema2013large, groenendijk2014photovoltaic,gabor2011hot,freitag2013photoconductivity}. In addition, the asymmetric tunnelling barrier is also known for generating a photocurrent without external bias \cite{yu2016unusually, hou2019photocurrent}. Here, we disentangle these electronic and thermal contributions and measure their corresponding response times by performing ultrafast photocurrent autocorrelation and non-degenerate pump-probe photocurrent measurements. We conclude that the photocurrent has approximately 2-ps fast dynamics for devices of various thicknesses and a slow dynamics that lasts for 25 ps or more.

Ultrafast optical techniques are powerful tools for probing the photocarrier dynamics \cite{hong2014ultrafast,jin2018ultrafast,daranciang2012ultrafast}. Photocurrent autocorrelation technique has been previously used to study the photocurrent dynamics in 2D-material based photodetectors \cite{Massicotte2016,massicotte2016photo,Sun2012,Graham2013,Betz2013}. In this approach, two equal-power laser pulses are split from a single femtosecond pulse and illuminate the heterostructure to generate a photocurrent (Fig. 1A). When these two pulses coincide on the device, a saturation in the photocurrent can be observed due to either electronic or thermal mechanisms. For example, electronic saturation may arise from absorption saturation and increased interactions between carriers, whereas thermal saturation may be induced by the nonlinear temperature increase under illumination \cite{nie2014ultrafast, vogt2020ultrafast, sun2014observation, wang2015ultrafast}. By scanning the delay between these two pulses, a characteristic time associated with the saturation mechanism can be measured. When multiple mechanisms contribute to the photocurrent, the intrinsic photocurrent time response can be obscured.
 
\section*{Results and Discussions}
We first perform the photocurrent autocorrelation measurement on a heterostructure composed of both bilayer (BL) and four-layer (4L) 3R-MoS\textsubscript{2}. The optical image of the heterostructure is shown in Fig. S1. From the current-voltage characteristics shown in Fig. 1B, a spontaneous photovoltaic behavior is clearly observed. The zero-bias photocurrent mapping shows a finite photoresponse in BL and 4-layer regions and the photocurrent distribution agrees with the MoS\textsubscript{2} thickness in the graphene overlapped area (Fig. 1C. Photocurrent maps for other devices can be found in Fig. S2). Such photocurrent grows sublinearly with optical power (Fig. S3), giving rise to a negative autocorrelation signal. A representative autocorrelation result is shown in Fig. 1d. A dip of the photocurrent strength is observed at zero delay, with symmetric recovery dynamics at both positive and negative delays, as expected for conventional photocurrent autocorrelation signals \cite{Massicotte2016,massicotte2016photo,Sun2012,Graham2013,Betz2013}.

More quantitative analysis of the autocorrelation signal in the bilayer area is presented in Fig. 2. With a fluence of \SI{140}{\micro\joule\per\square\centi\meter} for a spot size of approximately \SI{1}{\micro\meter} in diameter at 770 nm (50 fs pulse width) for each pulse, the saturation dip at zero time delay is approximately 15\% of the steady state signal. The recovery time is fitted to be 17 ps, more than one order of magnitude slower than the charge-transfer time at the MoS\textsubscript{2}/graphene interface \cite{yuan2018photocarrier, luo2021twist}, suggesting the saturation mechanism is likely not electronic. If such a slowing is caused by the charge transfer at the MoS\textsubscript{2}/MoS\textsubscript{2} interface, a previous study suggests an external electric field can speed up the process when the TMD is not monolayer \cite{Massicotte2016}. However, when we apply a bias voltage from -0.5 V to 0.5 V across the device, no significant change is observed in the recovery time (Fig. 2A inset. The raw data is shown in Fig. S4). 

Instead, we suggest that this photocurrent saturation originates from photo-thermal effects. As shown in Fig. 2C, the photoresponse of our device has a surprisingly strong temperature dependence. When the heterostructure is heated by 10 K above room temperature (RT), the photocurrent drops by more than $25\%$ (Fig. 2c inset). More strikingly, the photoresponsivity increases by more than one order of magnitude after cooling the heterostructure from RT to 3 K (Fig. 2C, see also in another device in Fig. S5, full temperature range in Fig. S6A). At the A-exciton resonance, the EQE approaches 10\%, which is remarkable for an atomically thin device. In addition, we also observe a significant temperature dependence in the total resistance of the device (Fig. S6B), which motivates us to develop a shunt-resistance model. In this model, within a few tens of Kelvin above room temperature, which is relevant to our pump-probe photocurrent measurement, only the shunt resistance is expected to change significantly due to its tunnelling current origin \cite{georgiou2013vertical}. By modeling the measured photocurrent to be proportional to the shunt resistance, which exponentially decreases with increasing temperature, we are able to fit the temperature dependence of the photocurrent (Fig. 2C inset, details in Supplementary Material). Other factors that can contribute to this strong temperature dependence include the variation in spontaneous polarization \cite{yasuda2021stacking, liu2022identifying}, and change in photocarrier recombination time or exciton linewidth, which affects the absorption of the heterostructure. As a result, when an ultrafast laser pulse is absorbed by the device, the transient temperature of the illuminated region quickly rises, which decreases the shunt resistance and subsequently decreases the photocurrent generated by the following pulse. In this case, the saturation recovery time corresponds to the device heat dissipation time. At low temperatures, the photocurrent becomes less temperature dependent (Fig. S6), which is consistent with the smaller autocorrelation signal observed at 3K, possibly due to the internal quantum efficiency limit.

We confirm the photo-thermal origin of the saturation by measuring the photocurrent autocorrelation at cryogenic temperature (Fig. 2B). After most phonon modes are frozen at low temperature, the thermal conductivity of the 2D materials significantly decreases \cite{Liu2017,Balandin2011}, which should lead to a longer heat dissipation time. Experimentally, we find a recovery time that is approximately three times longer (55 ps) than that at RT, when the heterostructure is cooled to 3 K. All these evidences suggest that the dynamics observed in the autocorrelation measurements is limited by the device heat dissipation.

In addition to this long recovery time, we also observe a transient middle peak in the autocorrelation measurement at 3 K. Exponential fitting of that small peak suggests a time constant of approximately 4 ps. Similar peaks have been observed in the 4L (Fig. S7A) region and on other bilayer devices of varying amplitudes up to RT (Fig. S7B). We interpret this feature as a result of the electronic temperature saturation in the graphene electrodes. As discussed in our previous work \cite{yang2022spontaneous}, the top and bottom graphene electrodes are naturally doped with charges of opposite polarities in these devices, which consequently acquire opposite Seebeck coefficients. Upon laser heating, a PTE current in a direction opposite to the photovoltaic-current is generated, which reduces the net photocurrent. In the autocorrelation experiment, the two subsequent laser pulses with a small delay time saturate the graphene electronic temperature, which decreases the PTE effect and therefore increases the overall photocurrent. The characteristic time of the middle peak also agrees with graphene electronic temperature relaxation through supercollision \cite{song2012disorder,Graham2013,Betz2013}.

Since the photo-thermal effect of the laser pulse obscures the electronic response speed in autocorrelation measurements, an alternative method is needed to separate the thermal and electronic contributions. Here we develop a non-degenerate pump-probe photocurrent spectroscopy technique that permits the determination of the intrinsic speed of our devices (Fig. 3A). First, we use a strong sub-bandgap infrared pulse (1030 nm, 50 fs, \SI{350}{\micro\joule\per\square\centi\meter}, IR-pulse) to heat the graphene electrodes, which subsequently heats the MoS\textsubscript{2}. Second, we employ a weak visible pulse (670 nm, 50 fs, \SI{2}{\micro\joule\per\square\centi\meter}, VIS-pulse) that is resonant with the A-exciton of MoS\textsubscript{2} to generate a photocurrent pulse but without generating much heat. 

At negative delays when the VIS-pulse arrives earlier than the IR-pulse, despite the fact that the IR-pulse generates a finite photocurrent, the photo-thermal saturation effect should be negligible because little heat is deposited by the VIS-pulse (details in the Supplementary Material). The photocurrent should decrease significantly immediately after the zero delay since the substantial heating induced by the IR-pulse would reduce the photocurrent generated by the VIS-pulse, in accordance with the results from our autocorrelation experiments. At a large positive delay, the photocurrent should be restored to the steady state as the heat is fully dissipated. Therefore, as we scan the time delay, we expect to see an asymmetric pump-probe photocurrent signal similar to photocarrier dynamics from an optical pump-probe measurement, which exhibits a quick change followed by an exponential decrease in signal. Since the pump-probe photocurrent dynamics is approximately a cross-correlation between the photocurrent and device temperature dynamics, the decay of the signal is likely limited by the heat dissipation, while the rise of the signal should be determined by the photocurrent decay and lattice heating.

Experimentally, a highly asymmetric temporal response is indeed observed in the pump-probe photocurrent measurement (Fig. 3B). The pump-probe photocurrent signal captures the photocurrent change due to thermal saturation. As expected, we find a very sharp drop of the photocurrent around the zero delay. The drop lasts for approximately 4 ps (Fig. 3B inset), which gives an upper bound for the characteristic time of the transient photocurrent. It is followed by a 20-ps slow recovery dynamics similar to that observed in the autocorrelation experiment, corresponding to the cooling of the device. Such pump-probe photocurrent results confirm the existence of a picosecond photocurrent response. In addition, a small photocurrent saturation is observed ahead of the sharp drop before zero delay, suggesting multiple dynamics could be associated with the photocurrent. The delayed IR-pulse can affect the photocurrent generated earlier by the VIS-pulse if the photocurrent lasts long enough to overlap with the heating pulse. With the fast photocurrent component determined from the quick drop, the saturation long before the zero delay indicates the existence of a slow photocurrent component. In addition, after symmetrizing the pump-probe photocurrent signal by adding itself to the time-reversed copy, the symmetrized result shows similar dynamics as the photocurrent autocorrelation signal, except that the autocorrelation is affected by the PTE effect near zero delay (details in the Supplementary Material). Since the signal in the pump-probe photocurrent experiment is much larger at positive delay than that at negative delay, we confirm the slow decay in the autocorrelation signal is dominated by the photo-thermal saturation effect.

To have a quantitative understanding of the photocurrent dynamics, a phenomenological model is developed to model the joint response of the device temperature evolution (thermal pulse) and transient photocurrent (photocurrent pulse). In this model, each VIS-pulse generates two concurrent photocurrent pulses in the device with independent decay constants $I(t)=I_1 e^{-t/t_1}+I_2 e^{-t/t_2}$. Upon overlapping with the thermal pulse, the instantaneous photocurrent is reduced to $\alpha I$, where $\alpha$ is a dimensionless saturation factor and its temperature dependence is defined in the Supplementary Material. As the device temperature is a function of time, $\alpha$ is consequently time-dependent. After calculating the evolution of the transient lattice temperature of the device excited by an IR-pulse at a delay of $\Delta \tau$ through a two-temperature model (Supplementary Material), we can obtain the average photocurrent through the cross-correlation relation $\bar I(\Delta \tau)=\frac{1}{\tau_{rep}}\int_{0}^{\tau_{rep}} \alpha(t)I(t+\Delta \tau) dt$, where $\bar I$ is the average photocurrent and $\tau_{rep}$ is the repetition time of the laser, which is much longer than the photocurrent response time. The calculated $\bar I$ fits our experimental results well (Fig. 3B) with the transient photocurrent profile $I(t)$ presented in Fig. 3C. The fast decay in the early stage has a time constant $t_1$ of approximately 2 ps, which is similar to the graphene/TMD charge-transfer time. The measured time constant is nearly independent of the power of the IR or VIS pulses. This characteristic time of the transient photocurrent approaches that of the charge transfer at the graphene-TMD interface \cite{hong2014ultrafast, yuan2018photocarrier, Zhang2017, luo2021twist,chen2019highly}, suggesting a likely faster interlayer charge-transfer process at the MoS\textsubscript{2} interface, similar as those observed in TMD heterostructures \cite{Hong2014}. On the other hand, the slow component has a time constant $t_2$ of approximately 25 ps. It was previously found that the two-component photocurrent can originate from defect-related processes \cite{wang2015ultrafast,vogt2020ultrafast,furchi2014mechanisms}, which agrees with the dynamics we observe.

To further confirm the electronic and thermal contributions in the photocurrent dynamics, a comparison is made between the BL and 4L regions with the pump-probe photocurrent measurements. Similar to the BL region, an asymmetric time dependence before and after zero-delay as well as a sharp drop in photocurrent around zero-delay are observed in the 4L (Fig. 4A, isolated curve in Fig. S8), indicating a similarly fast dynamics at the graphene/MoS\textsubscript{2} interface. Nevertheless, the pump-probe photocurrent signal before zero-delay is much more prominent in the 4L than BL, suggesting a much larger slow component contribution in the thick area. This contrast might result from a larger defect density in 2D or the extra MoS\textsubscript{2}-MoS\textsubscript{2} interfaces in the 4L. Interestingly, an external bias can be applied to tune the slow component of the photocurrent, which is observed before zero delay (Fig. 4A). As fitted by the aforementioned model, the slow component becomes faster when the electric field is along the photocurrent direction, and vice versa (Fig. 4B). On the other hand, the recovery time at positive delay is dominated by the device cooling dynamics, and is therefore largely bias independent. These results further confirm the thermal saturation nature and the existence of a two-component photocurrent response in the 3R-MoS\textsubscript{2} photovoltaic device. 

In conclusion, we have observed a large temperature dependence in the spontaneous photovoltaic effect in heterostructures comprising atomically thin 3R-MoS\textsubscript{2} and graphene. At low temperatures, the EQE can reach 10\% even in the thinnest area with only two atomic layers of MoS\textsubscript{2}. Upon laser heating, such a temperature dependence leads to a strong thermal saturation of the photocurrent, which dominates in the commonly used photocurrent autocorrelation measurement and obscures the photocurrent dynamics. In addition, we have developed a non-degenerate pump-probe photocurrent spectroscopy technique, which can distinguish the electronic and thermal dynamics. With this measurement technique, we find that the transient photocurrent has two contributions with distinct temporal responses. We quantify the fast photocurrent response to be on the picosecond level, which is similar to the charge-transfer process at the graphene-TMD interface, suggesting an intrinsic device bandwidth of hundreds of gigahertz. At room temperature, the photoresponsivity and speed of our homobilayer device are comparable to that of TMD heterobilayer devices with type-II band alignment \cite{lee2014atomically,Hong2014,ma2019recording}. Our results may stimulate future uses of ferroelectric van der Waals devices in optoelectronic applications that require high performance with low power consumption and built-in memory function, such as in high-speed optical communications and optical computing. Furthermore, the non-degenerate pump-probe photocurrent spectroscopy technique can also be applied to study the interplay between various photo-excitation mechanisms in other types of photosensors.

\section*{Materials and methods}
\textbf{Sample fabrication:} The 3R-$\text{MoS}_2$ flakes are exfoliated from a bulk crystal (HQ Graphene) on SiO\textsubscript{2}/Si substrates. Graphene electrodes are cut into specific shapes with a femtosecond laser. The flake thickness is identified by an optical microscope and reflection contrast spectroscopy. The 3R phase is confirmed by SHG measurement \cite{zhao2016atomically}. Exfoliated hexagonal boron nitride (hBN) is used for encapsulation. All devices are fabricated using dry transfer method under ambient conditions with a home-built transfer stage. Electrical contact is achieved by overlapping the graphene with gold electrodes pre-patterned by optical lithography on heavily p-doped Si/$\text{SiO}_2$ substrates.\\
\textbf{Photocurrent measurement:} For the autocorrelation measurements, a home-built wavelength-tunable optical parametric oscillator (OPO) (76MHz repetition rate, 50 fs pulse width) is used to illuminate the sample, which sits in a continuous-flow optical cryostat (Oxford Microstat-He), through a long working distance 100X objective lens (Mitutoyo, N.A. = 0.5). The signal is detected through a lock-in amplifier phase-locked to a mechanical chopper. In the pump-probe photocurrent measurement, we split some light from the pump laser (Light-conversion, 1030nm) of the OPO, and then combine it with the OPO beam for collinear measurements. The photocurrent is measured by modulating the VIS-pulse from the OPO. For wavelength-dependent measurements, a supercontinuum white light source (YSL)($\sim$100 ps pulse width, 10 MHz) is wavelength selected by an acousto-optic tunable filter to illuminate the device. For device photocurrent mapping, current-voltage characteristics and temperature dependence, a 532nm CW laser is used. 




\section*{Acknowledgments}
We thank Dr. Manabendra Kuiri and Ray Su for help on device fabrication, Dr. Jeff Young and Dr. Mengxing Na for help on measurements and modelling, respectively. \textbf{Funding:} We acknowledge support from the Natural Sciences and Engineering Research Council of Canada, Canada Foundation for Innovation, New Frontiers in Research Fund, Canada First Research Excellence Fund, and Max Planck–UBC–UTokyo Center for Quantum Materials. Z.Y. is also supported by the Canada Research Chairs Program. K.W. and T.T. acknowledge support from JSPS KAKENHI (Grant Numbers 19H05790, 20H00354 and 21H05233). \textbf{Author Contributions:} J.W., D.Y. and Z.Y. conceived this work. D.Y., J.L., and J.W. fabricated the samples. J.W. and D.Y. conducted the measurement with the help from J.D., Y.X. and J.L.. J.W., D.Y. and Z.Y. analyzed the data. M.W., E.O., and D.J. developed the optical parametric oscillator. K.W. and T.T. provided the hBN crystal. Z.Y. supervised the project. J.W., Z.Y. and J.D. wrote the manuscript based on the input from all other authors. J.W., D.Y. and J.L. contributed equally to this work. \textbf{Competing interests:} The authors declare no competing interests. \textbf{Data and materials availability:} All data needed to evaluate the conclusions in the paper are present in the paper and/or the Supplementary Materials.

\section*{Supplementary materials}
Fig. S1. Optical image of the device\\
Fig. S2. Photocurrent mapping of additional devices\\
Fig. S3. Illumination power dependence of photocurrent\\
Fig. S4. Bias-dependent auto-correlation measurements on the BL region\\
Fig. S5. Wavelength-dependent photoresponsivity of another BL device\\
Fig. S6. Temperature dependence of the photocurrent and device resistance\\
Fig. S7. Additional autocorrelation signals\\
Fig. S8. Pump-probe photocurrent measurement of the 4L region at zero bias \\
Note S1. Power dependence of the autocorrelation signal\\
Note S2. Circuit model\\
Note S3. Temperature dependence of the shunt resistance\\
Note S4. Photocurrent temperature dependence\\
Note S5. IR-pulse induced photocurrent\\
Note S6. Symmetrized pump-probe signal and its comparison with the autocorrelation result\\
Note S7. Two-temperature model\\
Fig. S9. Power-dependent autocorrelation measurements of BL region\\
Fig. S10. Circuit model of the Gr/3R-MoS\textsubscript{2}/Gr photodetector\\
Fig. S11. Band alignment of Gr/3R-MoS\textsubscript{2}/Gr junction\\
Fig. S12. Photocurrent power dependence under IR-pulse excitation\\
Fig. S13. Comparison of symmetrized pump-probe and autocorrelation photocurrent signals\\
Fig. S14. Graphene electronic heat capacity\\



\clearpage

\begin{figure}[h]%
\makebox[\textwidth][c]{%
\includegraphics[width=1\textwidth]{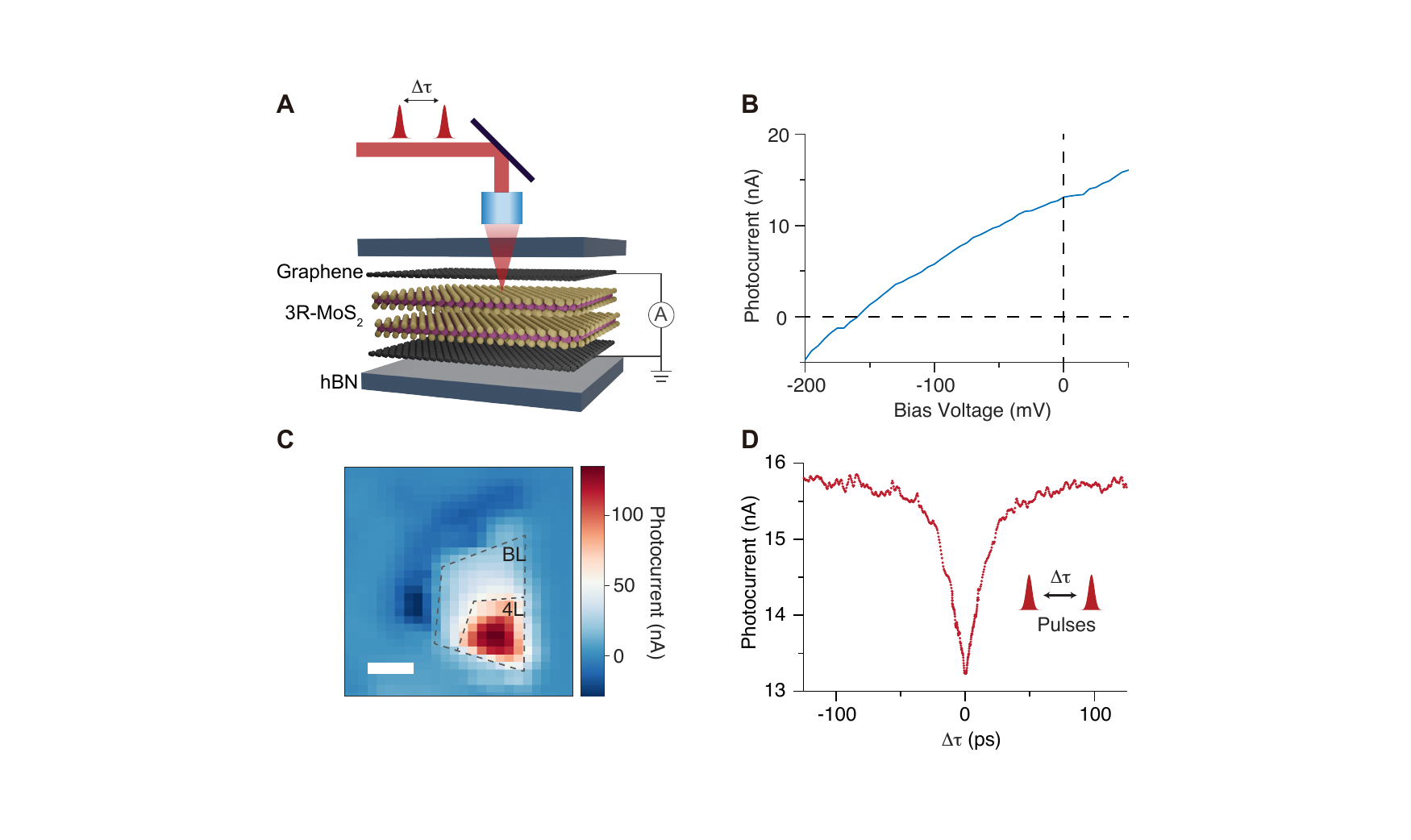}
}
\noindent{\textbf{Fig. 1. Photocurrent generation in a Gr/3R-MoS\textsubscript{2}/Gr device.} (\textbf{A}) A schematic of the device and ultrafast time-resolved photocurrent measurement. The heterostructure consists of a bilayer or few-layer 3R-MoS\textsubscript{2} flake that is sandwiched between two graphene electrodes. The whole device is encapsulated in hBN. The photocurrent is measured with a lock-in amplifier through graphene electrodes. The two ultrafast pulses are seperated by a time delay $\Delta \tau$, and focused collinearly on the device. (\textbf{B}) Photocurrent-voltage characteristics on a bilayer device at room temperature. The device shows photovoltaic effect with nonzero photocurrent without bias. (\textbf{C}) Photocurrent map of a device consists of both bilayer and 4-layer regions at zero bias. The resolution of the scan is 400 nm in both directions. The darker blue regions are also bilayer but with a different polarity. Scale bar: \SI{2}{\micro\metre}. Both (\textbf{B}) and (\textbf{C}) are measured under 20 uW 532 nm CW laser illumination. (\textbf{D}) Autocorrelation signal from the BL and 4L region with 770 nm femtosecond pulse excitation, which is above the indirect bandgap of the material. }
\label{fig1}
\end{figure}

\begin{figure}[h]%
\makebox[\textwidth][c]{%
\includegraphics[width=0.5\textwidth]{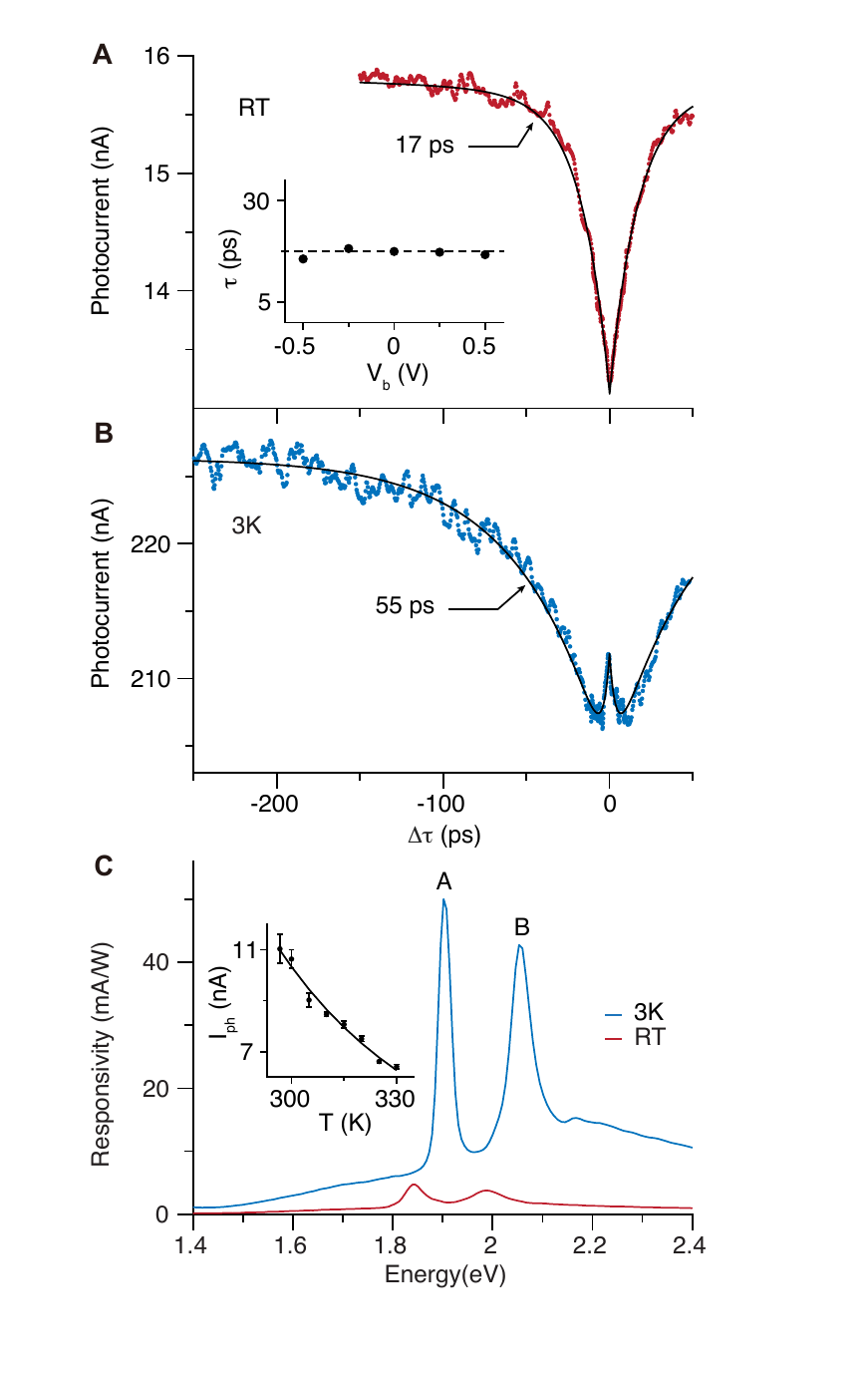}
}
\noindent{\textbf{Fig. 2. Temperature-dependent measurements in the BL region.} Autocorrelation signals of the bilayer region measured at (\textbf{A}) room temperature and (\textbf{B}) 3K. The dotted lines are from experimental measurements, and black lines are exponential fittings to extract the time constants. Inset of (\textbf{A}): photocurrent time constant extracted from bias-voltage-dependent autocorrelation measurements. The dashed line indicates a photocurrent time constant, $\tau$, of 17 ps. (\textbf{C}) Energy-dependent photoresponsivity of the bilayer region measured at room temperature and 3K. The A and B peaks correspond to the A-exciton and B-exciton absorptions in bilayer 3R-MoS\textsubscript{2}. The average power is kept at around \SI{0.5}{\micro\watt} at all wavelengths to avoid saturation. Inset: photocurrent of the BL region at temperatures above RT with 532 nm CW laser excitation. The dots are experimental data and the solid line is fitted based on the shunt-resistance model. }
\label{fig2}
\end{figure}

\begin{figure}[h]%
\makebox[\textwidth][c]{%
\includegraphics[width=1\textwidth]{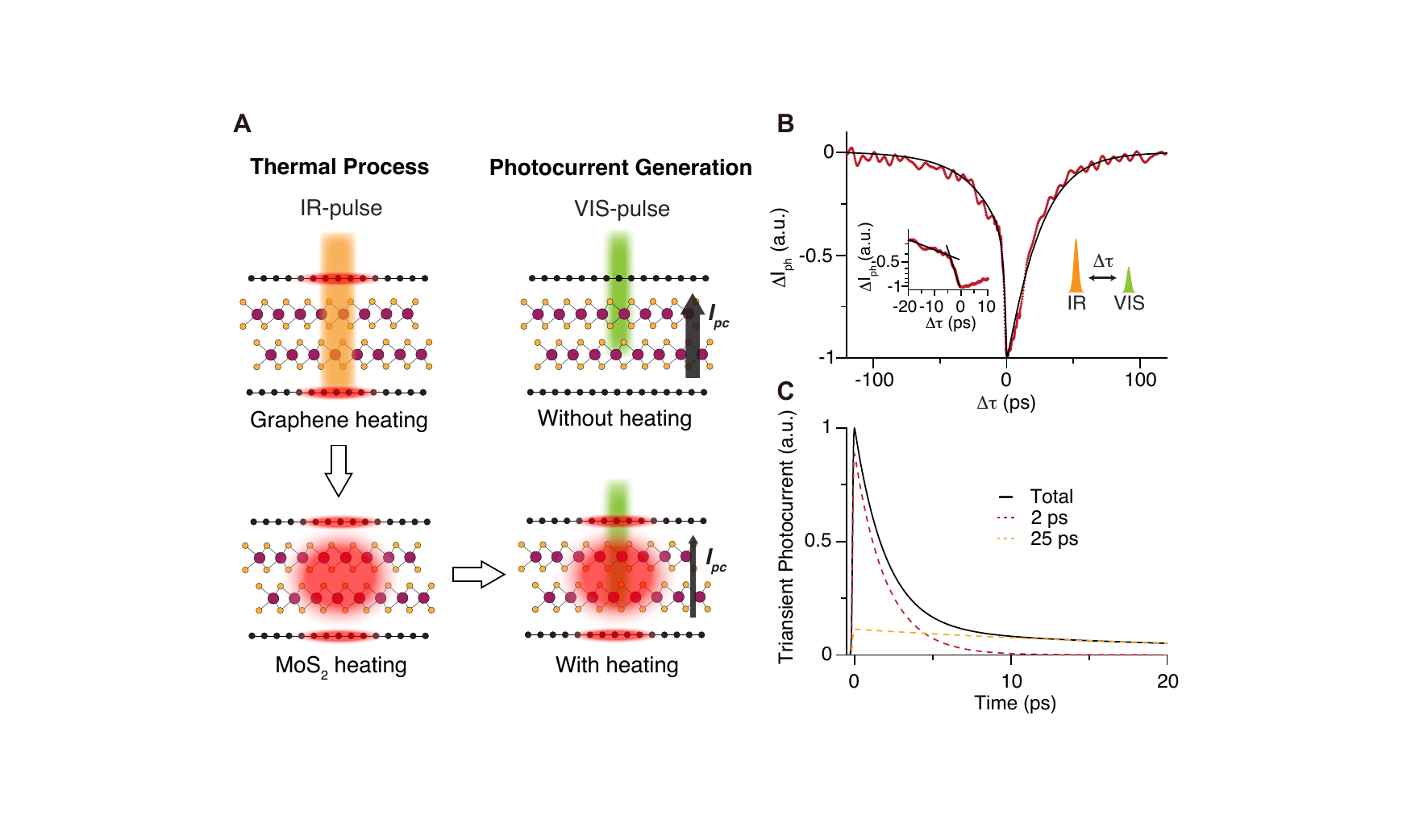}
}
\noindent{\textbf{Fig. 3. Pump-probe photocurrent measurement.} (\textbf{A}) Illustration of the heating and cooling dynamics of the device by a sub-bandgap pulse (IR-pulse). Graphene is firstly heated by the IR-pulse and then heats MoS\textsubscript{2}. The photocurrent is generated by the VIS-pulse, and it gets smaller when the device is thermally saturated upon heating. (\textbf{B}) Pump-probe photocurrent measurement on a bilayer device. The red dotted line represents experimental results and the solid black curve is the fitting from the numerical model. The measurements are carried out with an IR-pulse(VIS-pulse) fluence of {\SI{350(2)}{\micro\joule\per\square\centi\meter}}. The pump-probe photocurrent signal is normalized for comparison with the numerical model. The 20 ps scan around zero delay is performed at a finer temporal resolution than the rest. Inset on the left: A zoom-in semilog plot of the pump-probe photocurrent signal near zero delay. The black straight lines are for guidance. (\textbf{C}) Transient photocurrent responses for different processes. The total profile consists of two photocurrent generation processes, with the fast and slow processes having characteristic times of 2 ps and 25 ps, respectively.}\label{fig3}
\end{figure}

\begin{figure}[h]%
\makebox[\textwidth][c]{%
\includegraphics[width=0.5\textwidth]{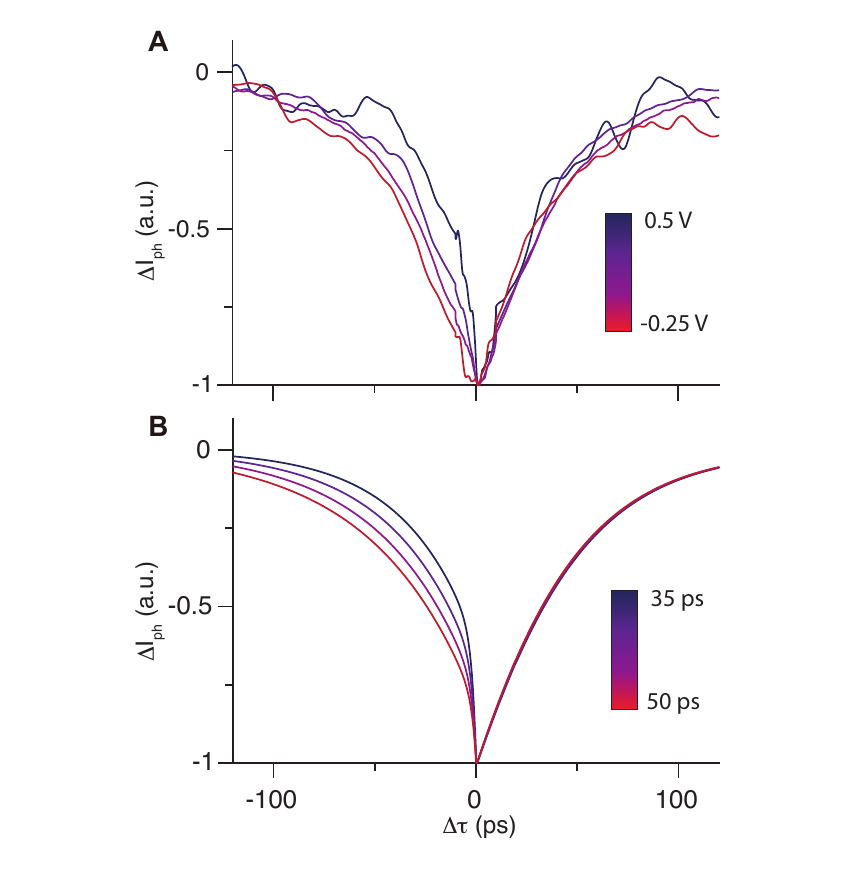}
}
\noindent{\textbf{Fig. 4. Pump-probe photocurrent measurement in the 4-layer region.} (\textbf{A}) Bias-dependent pump-probe photocurrent measurement in the 4L region with bias voltages ranging from -0.25 V to 0.5 V. The measurements are carried out with an IR-pulse(VIS-pulse) fluence of approximately \SI{350(2)}{\micro\joule\per\square\centi\meter}.  (\textbf{B}) Numerical modeling results with the response time of the slow component varying from 35 to 50 ps.}\label{fig4}
\end{figure}

\end{document}